\documentclass{emulateapj}

\newcommand{\myemail}{sascha.quanz@astro.phys.ethz.ch}

\slugcomment{To appear in ApJ Letters}

\shorttitle{Gaps in the HD169142 protoplanetary disks revealed by polarimetric imaging}
\shortauthors{Quanz et al.}

\begin{document}


\title{Gaps in the HD169142 protoplanetary disk revealed by polarimetric imaging:  Signs of ongoing planet formation?}

\author{Sascha P. Quanz$^{1,2}$, Henning Avenhaus$^2$, Esther Buenzli$^3$, Antonio Garufi$^2$, Hans Martin Schmid$^2$, and Sebastian Wolf$^4$}
\email{\myemail}

\altaffiltext{1}{Based on observations collected at the European Organisation for Astronomical Research in the Southern Hemisphere, Chile, under program number 089.C-0611(A).}
\altaffiltext{2}{Institute for Astronomy, ETH Zurich, Wolfgang-Pauli-Strasse 27, 8093 Zurich, Switzerland}
\altaffiltext{3}{Department of Astronomy and Steward Observatory, University of Arizona, Tucson, AZ 85721, USA}
\altaffiltext{4}{University of Kiel, Institute of Theoretical Physics and Astrophysics, Leibnizstrasse 15, 24098 Kiel, Germany}

\begin{abstract}
We present $H$-band VLT/NACO polarized light images of the Herbig Ae/Be star HD169142 probing its protoplanetary disk as close as $\sim$0.1$''$ to the star. Our images trace the face-on disk out to $\sim$1.7$''$ ($\sim$250 AU) and reveal distinct sub-structures for the first time: 1) the inner disk ($\lesssim$20 AU) appears to be depleted in scattering dust grains; 2) an unresolved disk rim is imaged at $\sim$25 AU; 3) an annular gap extends from $\sim$40 -- 70 AU; 4) local brightness asymmetries are found on opposite sides of the annular gap. We discuss different explanations for the observed morphology among which ongoing planet formation is a tempting -- but yet to be proven -- one. Outside of $\sim$85 AU the surface brightness drops off roughly $\propto r^{-3.3}$, but describing the disk regions between 85--120 AU / 120--250 AU separately with power-laws $\propto r^{-2.6}$/$\propto r^{-3.9}$ provides a better fit hinting towards another discontinuity in the disk surface.  The flux ratio between the disk integrated polarized light and the central star is $\sim 4.1\cdot 10^{-3}$. Finally, combining our results with those from the literature, $\sim$40\% of the scattered light in the $H$-band appears to be polarized. Our results emphasize that HD169142 is an interesting system for future planet formation or disk evolution studies. 
\end{abstract}



\objectname{HD169142} 


\section{Introduction}
To study the physical and chemical conditions for planet formation and to search for morphological evidence of forming planets, the inner few tens of AU of protoplanetary disks have to be investigated. In particular, annular gaps in those disks are considered to be possible signatures of ongoing planet formation. In this letter we report the detection of such gaps in the disk around the Herbig Ae/Be star HD169142 using polarimetric differential imaging (PDI). PDI is a powerful high-contrast technique to study the dusty surface layer of protoplanetary disks very close to the star \citep[e.g.,][]{quanz2011b,hashimoto2011,muto2012,quanz2012a,hashimoto2012,tanii2012,kusakabe2012,mayama2012,grady2013}.  

\begin{deluxetable}{llc}
\centering
\tablecaption{Basic parameters of HD169142. 
\label{parameters}}           
\tablehead{
\colhead{Parameter} & \colhead{Value for HD169142} & \colhead{Reference\tablenotemark{a}}
}
\startdata
RA (J2000) & 18$^h$24$^m$29$^s$.79  & (1) \\ 
DEC (J2000) & -29$^\circ$46$'$49$''$.22   & (1)\\
$J$ [mag] & 7.31$\pm 0.02$  & (1)\\
$H$ [mag]& 6.91$\pm 0.04$   &(1)\\
$K_s$ [mag]& 6.41$\pm 0.02$  & (1)\\
Sp. Type & A9III/IVe / A7V& (2),(3) \\
$v\,{\rm sin}\,i$ [km$\,$s$^{-1}$] & 55$\pm 5$ & (2)\\
Age [Myr]& 1--5 / 12  / 3--12  & (2),(3),(4) \\
$[{\rm Fe/H}]$ & $ -0.5\pm 0.1$ / -0.25 -- -0.5 & (2),(5) \\
log $g$ & 3.7$\pm$0.1 / 4.0--4.1 & (2),(5)\\
$T_{\rm eff}$ [K] & 7500$\pm$200 / 6500 / 7650$\pm$150 & (2),(3),(5)\\
Mass [M$_\sun$]  & $\sim$1.65 & (3)\\ 
$R_*$  [R$_\sun$]& $\sim$1.6 & (3),(5) \\
$L_*$ [L$_\sun$] & $\sim$8.6  & (3) \\
$\dot{M}$ [$10^{-9}$ ${\rm M}_\sun{\rm yr}^{-1}$]& $\sim$3.1 / $\leq$1.25$\pm$0.55  & (3),(4) \\
Distance [pc] & 145\tablenotemark{b} / 151& (6),(3)\\
\enddata
\tablenotetext{a}{References --- (1) 2MASS point source catalog \citep{cutri2003}, (2) \citet{guimaraes2006}, (3) \citet{blondel2006}, (4) \citet{grady2007}, (5) \citet{meeus2010}, (6) \citet{sylvester1996}.}
\tablenotetext{b}{adopted value in this paper}
\end{deluxetable}

Our new $H$-band PDI observations of HD169142 resolve for the first time distinct structures in the inner disk regions with high signal-to-noise and allow an in-depth analysis of the disk profile and morphology. The basic stellar parameters of HD169142 are summarized in Table~\ref{parameters}. 

\begin{deluxetable*}{lccccc}
\centering
\tablecaption{Summary of NACO/PDI observations of HD169142 on July 25, 2012.
\label{observations}}           
\tablehead{
\colhead{Filter} & \colhead{DIT $\times$ NDIT\tablenotemark{a}} & \colhead{Dither}  & \colhead{Average}  &  \colhead{Average} & \colhead{Average $\langle \tau_0 \rangle$\tablenotemark{d} }\\
\colhead{} & \colhead{} & \colhead{positions\tablenotemark{b}}  & \colhead{airmass}  &\colhead{seeing\tablenotemark{c}} & \colhead{}
}
\startdata
$NB1.64$ & 1 s $\times$ 15 & 3 & $\sim$1.02  & $\sim 1.57''$ & $\sim 14 $ ms\\
$H$ & 1 s $\times$ 45 & 2$\times$6 &  $\sim$1.06  & $\sim 1.04''$ &$\sim 22 $ ms \\
$NB1.64$ & 1 s $\times$ 20 & 3  & $\sim$1.11 & $\sim 1.07''$ & $\sim 20 $ ms\\

\enddata
\tablenotetext{a}{Detector integration time (DIT) $\times$ number of integrations (NDIT), i.e., total integration time per dither position and per retarder plate position.}
\tablenotetext{b}{At each dither position 4 different HWP positions (0.0$^\circ$, -22.5$^\circ$, -45.0$^\circ$, -67.5$^\circ$) were used.}
\tablenotetext{c}{DIMM seeing in the optical as measured from seeing monitor .}
\tablenotetext{d}{Average coherence time of the atmosphere calculated by the Real Time Computer of the AO system.}
\end{deluxetable*}

HD169142 was classified as a Herbig group Ib member \citep{meeus2001}, indicating that its spectral energy distribution (SED) rises in the mid-infrared (MIR), but does not show a 10 $\mu$m silicate emission feature \citep[e.g.,][]{boekel2005}. However, PAH emission features have been detected in MIR spectroscopy \citep{meeus2001,boekel2005} and were spatially resolved at 3.3 $\mu$m \citep{habart2006}. The disk was previously resolved in polarized light in the NIR from the ground \citep{kuhn2001,hales2006}, but those data did not allow an in-depth analysis of the disk structure and profile. \citet{grady2007} and \citet{fukagawa2010} detected the disk in scattered light at 1.1 $\mu$m with HST/NICMOS coronagraphy and in $H$-band from the ground, respectively, and they derived azimuthally averaged surface brightness profiles. In all direct imaging studies the disk was consistent with being seen face-on and no sub-structures were detected. In the MIR the disk was resolved at 12, 18, 18.8 and 24.5 $\mu$m \citep{honda2012, marinas2011} and most of this emission is thought to arise from disk regions between $\sim$30 -- 60 AU. From dust continuum and molecular line observations the mass and outer radius of the disk are estimated to be $\sim$0.005 - 0.04 M$_\sun$ and $\sim$235 AU \citep{raman2006, dent2006, panic2008,meeus2010,sandell2011}. The kinematic pattern and molecular line profiles are best fitted with a disk inclination of $\sim$13$^\circ$ \citep{raman2006,panic2008}. The lack of a silicate emission feature at 10 $\mu$m can be explained with a lack of small ($<$3--5 $\mu$m) dust grains in the inner $\sim$10 -- 20 AU of the disk possibly due to grain growth \citep{boekel2005}. From SED modeling a disk hole was inferred for the inner disk regions \citep{grady2007, meeus2010, honda2012}.

\begin{figure*}
\centering
\plotone{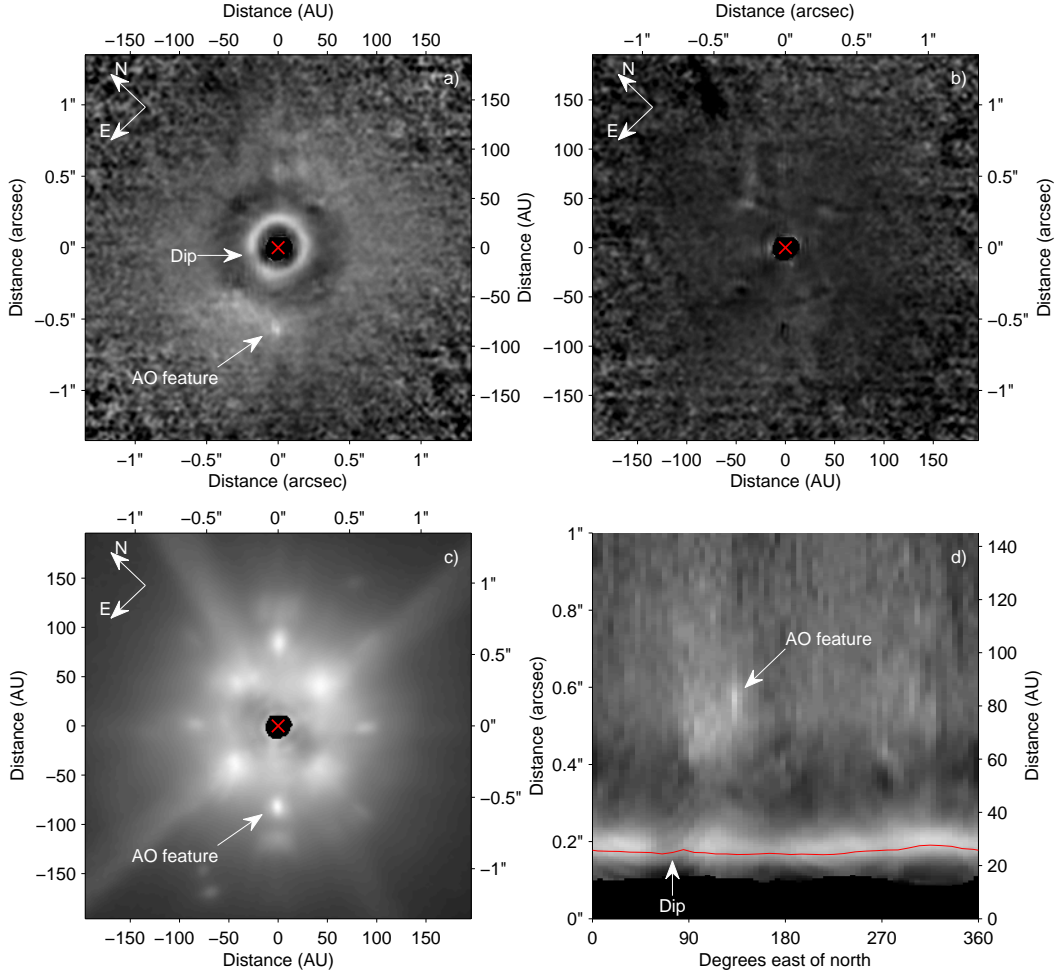}
\caption{NACO/PDI observations of HD169142 in  the $H$ band. a) Final $Q_{\rm r}$ image scaled with $r^2$ to compensate for the decrease in stellar flux (image shown in a linear stretch). The position of the central star is indicated by the red cross. Saturated pixels in the central regions have been masked out. Our data reveal a bright inner ring, a large gap and a smooth outer disk in polarized light. A brightness dip in the ring and a residual AO feature are indicated by arrows. b) $U_{\rm r}$ with the same scaling and stretch as for the $Q_{\rm r}$ image. c) Intensity image also scaled with $r^2$. Features from the AO system and the telescope spiders are clearly seen. d) Polar coordinate mapping of $Q_{\rm r}$. The innermost, masked out region is less than 0.1$''$ in diameter. The red line traces the peak brightness of the inner ring.
\label{images}}
\end{figure*}

\section{Observations and data reduction}\label{observations_section}
The observations were carried out with VLT/NACO \citep{lenzen2003,rousset2003} in the $H$ and $NB1.64$ filter. 
A description of the camera properties and the PDI observing mode is given in \citet{quanz2011b}. In short, a Wollaston prism splits the light in an ordinary and extraordinary beam with orthogonal, linear polarization directions. Both beams are imaged simultaneously on the detector. The polarization direction can be changed by rotating a half-wave plate (HWP). One full polarization cycle includes exposures at four HWP positions (0$^\circ$ and $-$45$^\circ$ for Stokes $Q$, $-$22.5$^\circ$ and $-$67.5$^\circ$ for Stokes $U$). We summarize the observations for HD169142 in Table~\ref{observations}. The central few pixels of the point spread function (PSF) were saturated in the $H$ filter. The exposures in the $NB1.64$ filter were unsaturated and used for photometric calibration (see below).

NACO suffers from instrumental polarization and cross-talk between the different Stokes components \citep{witzel2011,quanz2011b}. To obtain reliable results in PDI observations these effects need to be corrected for. Having obtained data sets for five more Herbig Ae/Be stars allowed us to do a systematic analysis and we refer to an upcoming publication for a detailed description of our data reduction and calibration approach (Avenhaus et al., in prep.). The key steps are:

All exposures were dark current corrected and flat-fielded. Bad pixels were replaced with the mean value of surrounding pixels. Each quadrant of the detector showed a noise pattern affecting every other row, which was eliminated by subtracting the mean value of each row from each pixel in that row. Detector regions close to the star were excluded to compute the mean values. The ordinary and extraordinary images in each exposure were extracted and the stellar position in all individual images determined. We rebinned the images to a factor three higher resolution (bicubic interpolation) and shifted all images to a common center (bilinear interpolation). Assuming that the central star is unpolarized in $H$-band we corrected instrumental polarization and polarization cross-talk effects\footnote{The results were consistent with those using the approach described in \citet{quanz2011b}.}. For each dither position the fractional Stokes parameters $p_Q$ and $p_U$ were computed using the double ratio approach \citep[see e.g.,][]{tinbergen1996,schmid2006}:
\begin{equation}
p_Q=\frac{{\rm R}_Q-1}{{\rm R}_Q+1}\quad;\quad p_U=\frac{{\rm R}_U-1}{{\rm R}_U+1}
\end{equation}
with
\begin{equation}
{\rm R}_Q=\sqrt{\frac{{I^{0^{\circ}}_{\rm ord}}/{I^{0^{\circ}}_{\rm extra}}}{I^{-45^{\circ}}_{\rm ord}/I^{-45^{\circ}}_{\rm extra}}}\quad ; \quad {\rm R}_U=\sqrt{\frac{{I^{-22.5^{\circ}}_{\rm ord}}/{I^{-22.5^{\circ}}_{\rm extra}}}{I^{-67.5^{\circ}}_{\rm ord}/I^{-67.5^{\circ}}_{\rm extra}}}
\end{equation}
Here, the subscripts refer to either the ordinary or extraordinary beam and the superscripts refer to the angular position of the HWP. The final $p_Q$ and $p_U$ images were obtained by averaging over all dither positions. From  $p_Q$ and $p_U$ and the intensity images $I$ we obtained the Stokes $Q$ and $U$ images, from which we computed the radial Stokes parameters $Q_{\rm r}$ and $U_{\rm r}$:
\begin{equation}
Q_{\rm r}=+Q\,{\rm cos}\,2\phi+U\,{\rm sin}\,2\phi
\end{equation}
\begin{equation}
U_{\rm r}=-Q\,{\rm sin}\,2\phi+U\,{\rm cos}\,2\phi
\end{equation}
with
\begin{equation}
\phi ={\rm arctan} \frac{x-x_0}{y-y_0}+\theta
\end{equation}
being the polar angle of a given position $(x, y)$ on the detector, $(x_0,y_0)$ denoting the central position of the star \citep[see, e.g.,][]{schmid2006} and $\theta$ being an offset due to polarization cross-talk effects (Avenhaus et al., in prep.), which amounts to 7$^\circ$ in our case. $Q_{\rm r}$ is equivalent to the polarized flux $P$ under the assumption that the polarized flux has only a tangential component, but, in contrast to $P$, $Q_{\rm r}$ is free of any biases introduced from computing the squares of the $Q$ and $U$ components \citep{schmid2006}. Under the same assumption, and assuming the correction for instrumental effects was done properly, $U_{\rm r}$ only contains noise and can be used to estimate the error in $Q_{\rm r}$ via
\begin{equation}
\Delta Q_{\rm r} = \sqrt{\sigma^2_{U_{\rm r}}} / \sqrt{n_{\rm res}}\quad.
\end{equation}
Here, $\Delta Q_{\rm r}$ refers to the 1$\sigma$ uncertainty in $Q_{\rm r}$, $\sigma^2_{U_{\rm r}}$ is the variance in the final $U_{\rm r}$ image and $n_{\rm res}$ the number of resolution elements in the region of interest. 

The calibration of the surface brightness level in the final $Q_{\rm r}$ image was done using the non-saturated images in the $NB1.64$ filter following the description in \citet{quanz2011b}. We estimate that the absolute flux calibration is good to $\sim$30\%.

\section{Results and analysis}\label{results}
Figure~\ref{images} shows the final $Q_{\rm r}$ and $U_{\rm r}$ images, an intensity image, and a polar coordinate mapping of the $Q_{\rm r}$ image. While the $Q_{\rm r}$ image shows extended polarized flux that we interpret as the disk around HD169142, the $U_{\rm r}$ image does not reveal any significant signal as expected for scattering of dust particles without any preferred alignment. 

Figure~\ref{images} a) reveals a protoplanetary disk with a complex radial morphology. Very close to the star, but outside the innermost, saturated pixels, the detected polarized flux is low and it increases going outwards to reach a ring-like maximum at $\sim$0.17$''$ ($\sim$25 AU). Outside of the narrow ring the surface brightness decreases and an annular  gap stretches from $\sim$0.28--0.48$''$ ($\sim$40 -- 70 AU) with a local minimum at $\sim$0.38$''$ ($\sim$55 AU). Outside of $\sim$0.52$''$ ($\sim$75 AU) the disk surface brightness drops off smoothly.

The bright inner ring does not appear fully circular (Figure~\ref{images} panels a and d) and the radius of the peak brightness varies between $\sim$0.16$''$ and $\sim$0.19$''$ with a mean value of $\sim$0.17$''$. A dip in the brightness of the ring seems to be present at position angle (PA) $\approx$80$^\circ$ (east of north). On the opposite side of the annular gap at PA$\approx$90$^\circ$, next to an AO artifact, a region of enhanced brightness is seen stretching a bit into the gap. The full-width-half-maximum of the ring varies slightly with PA but the ring is not spatially resolved. Using the ring we estimated an upper limit in disk inclination of $i<20^\circ$, so that $i$ is still in agreement with the fluctuations in ring radius. However, our image is consistent with a face-on disk. Analyzing the mean surface brightness as a function of azimuth for different radii also supports a small inclination angle as no significant variation is seen (\citet{muto2012} used this method for the SAO206462 disk).

\begin{figure}
\centering
\epsscale{1.2}
\plotone{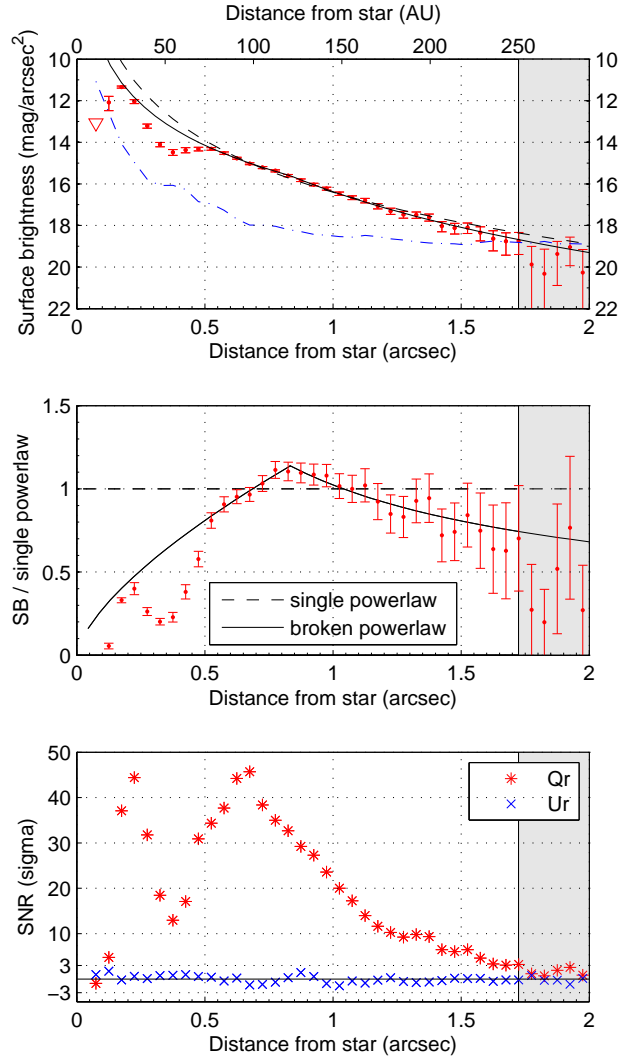}
\caption{Azimuthally averaged surface brightness profile and SNR for the HD169142 disk observed with NACO/PDI in the $H$ band. Top: Surface brightness in polarized flux. Error bars are 1$\sigma$ and include the the error defined in eq. (6) and the standard deviation from the mean flux in each annulus. The triangle represents a 1$\sigma$ upper limit. In the gray shaded region the noise dominates. The blue dash-dotted curve shows 3$\sigma$ the detection limit. The black dashed line is a power-law fit to the brightness profile between 85--250 AU. The black solid line is a broken power-law with a break point at 120 AU (see text). Middle: The observed surface brightness profile (red data points) and both power-law fits (dashed line = single power-law; broken power-law = solid line) divided by the single power-law fit. A broken power-law clearly provides a better fit to the data. Bottom: SNR for the $Q_{\rm r}$ (red stars) and $U_{\rm r}$ (blue crosses) images. The signal in $U_{\rm r}$ is consistent with random noise, while in  $Q_{\rm r}$ the disk can clearly be traced out to $\sim$1.7$''$.
\label{profile}}
\end{figure}

In Figure~\ref{profile} we show the azimuthally averaged surface brightness profile and the signal-to-noise profile for the $Q_{\rm r}$ and $U_{\rm r}$ images. Fitting a power-law to the brightness profile between 85--250 AU yields an exponent of $-3.31\pm0.11$, where the error denotes the 95\% confidence level (black dashed line in top panel of Figure~\ref{profile}). However, fitting the regions between 85--120 AU and 120--250 AU separately with power-laws yields a mathematically better fit in a $\chi^2$ sense (black solid line). In this case the power-law exponents are $-2.64^{+0.15}_{-0.17}$ and $-3.90^{+0.10}_{-0.11}$ for the inner and outer region, respectively. The power-law fit to the inner  region connects well with the bright inner ring at $\sim$25 AU.  A 0.15$''\times$0.1$''$ region centered on the AO feature seen in Figure~\ref{images} has been masked out before the power-law fits were computed. 

The maximum surface brightness in polarized flux of $S \approx$ 11.3 mag/arcsec$^2$ is reached at the location of the bright ring. Close to the inner edge of the outer disk ($\sim$75 AU) we find  $\approx$14.3 mag/arcsec$^2$ and at $\sim$220 AU $\approx$18.1 mag/arcsec$^2$. Inside the gap ($\sim$55 AU) the surface brightness is reduced by a factor of $\sim$3 compared to the corresponding value of the power-law fit. Integrating the polarized surface brightness between $\sim$14.5--250 AU and comparing it to the $H$-band flux of the central star yields a flux ratio of $F_{\rm Disk}^{\rm polarized}/F_{\rm Star}\approx4.1\cdot10^{-3}$. Here, the regions between $\sim$14.5--40 AU (the ring), $\sim$40--70 AU (the gap), and $\sim$70--250 AU (the outer disk) contribute $\sim$49\%, $\sim$14\%, and $\sim$37\% to the integrated disk flux, respectively.

\section{Discussion}
The face-on orientation and general extent of the disk have been reported in earlier scattered or polarized light studies \citep{kuhn2001,hales2006, grady2007, fukagawa2010}. However, no sub-structures were revealed so far. 

\subsection{The bright inner ring}
We interpret the bright ring as the inner rim of the main disk. Its location fits well to recent model predictions \citep{meeus2010,honda2012} but is not in agreement with a rim at 44 AU as suggested by \citet{grady2007}. The drop in polarized light inside of the rim is significant and supports the idea that a hole might be present in the inner $\lesssim$20 AU. The brightness dip in the rim at PA$\approx$80$^\circ$ hints to additional substructure or local change in dust grain properties. Although local minima in polarized flux can result from the scattering function of the dust grains even for small disk inclinations \citep[see, e.g.,][]{perrin2009,muto2012} also the outer disk regions should show a minimum at the same PA, which is not the case.

\subsection{The annular gap}
To our knowledge this is the first time that such a  symmetric and finite gap-like structure has been detected on the surface layer of a protoplanetary disk. 
This gap does not appear to be completely void of scattering material as the remaining surface brightness is above our detection limits. Seeing a gap-like feature in scattered light could indicate a deficiency in surface density possibly penetrating deeper into the disk mid-plane. If so, one (or several) forming planet(s) could be responsible for the observed morphology. Theory and simulations predict that planets open annular gaps in disks and the gap width and depth depend on the planet mass and on local disk properties \citep[e.g.,][]{lin1986,bryden1999,ruge2013}. \citet{jang-condell2012} simulated how planet-induced gaps would alter scattered light images of disks. They showed that depending on the mass of the planet the gap would not necessarily appear empty \citep[cf.][]{dong2012}. In addition, SPH simulations suggest that localized streams through the disk gap connect the outer and inner disk with the forming planet \citep[e.g.,][]{ruge2013}. One could speculate whether the observed local brightness asymmetries on opposite sides of the gap (see section 3) could be signposts of such features.  Finally, according to the simulations by \citet{jang-condell2012}, the impact of gaps and planets on the SEDs of star+disk systems is limited. This could explain why previous SED modeling attempts for HD169142 did not reveal this feature. 

One could in principle infer mass constraints for the companion from the geometry of the gap \citep{lin1986,bryden1999}. However, this exercise requires some specific assumptions about the protoplanetary disk (e.g., viscosity, scale height) which are poorly or not at all constrained empirically. Additionally, not knowing how empty the gap is in the disk mid-plane complicates any attempt to derive a useful companion mass estimate analytically and we leave it out of the scope of this letter.  However, a direct search for companions could be attempted with high-contrast imaging instruments even though this might be challenging if the planet is still embedded \citep{wolf2005,wolf2007}.

We strongly emphasize that there are alternative explanations for the observed disk morphology without invoking an embedded planet. For instance, the disk rim could cast some shadow onto the disk surface. Puffed-up disk rims have been predicted at least for the inner edges of 'typical' protoplanetary disks even if the exact geometry and the question whether these rims can cause disk shadows is still debated \citep[see, ][for a review]{dullemond2010}. Concerning HD169142, the disk model by \citet{honda2012} does \emph{not} predict a shadowing disk rim. However, a combination of radially varying dust grain properties with dust settling might possibly be able to create a shadow-like feature without the need of a local deficiency in the surface density. Dust continuum observations with sufficient spatial resolution can shed light on this open question.

\subsection{The disk surface brightness}
The single power-law index we derive from fitting the surface brightness profile agrees with the results from \citet{fukagawa2010}, who found  $S \propto r^{-3.4\pm0.4}$ between 120 -- 200 AU in $H$ band scattered light images. \citet{grady2007} found a slightly shallower profile with  $S \propto r^{-3.0\pm0.1}$ between $\sim$70 -- 200 AU using HST/NICMOS at 1.1$\mu$m. 

The errors in our profile fits justify the use of a broken power law. Broken power-law surface brightness profiles are observed in resolved scattered light images of debris disks, where the inner disk regions show a shallower profile due to the presence of larger grains while the outer disk profile is steeper and dominated by blown-out smaller grains \citep[e.g.,][]{augereau2001b,liu2004}. The disk around HD169142 is certainly still gas rich, but the observed brightness profile could be indicative of different dust grain populations. However, also changes in the dust density distribution could lead to the observed effect. Until now, all studies focusing on the gas content or the dust continuum of the HD169142 disk did not have enough spatial resolution to securely detect gaps or discontinuities in the disk structure. Of particular interest would be to see whether the broken power law discovered here is accompanied with a radial change in the dust-to-gas ratio.

Comparing the disk surface brightness derived by \citet{fukagawa2010} with our values between 120 and 200 AU suggests that roughly $\sim$40\% of the scattered light of HD169142 seems to be polarized in the $H$ band. However, as the errors bars in both studies are large, this value for the polarization fraction should be taken as a rough indication. 
At this level the polarization fraction appears to be higher than that for the HD100546 disk \citep[$\sim$14\%;][]{quanz2011b} but comparable to the value for the transition disk around AB Aur \citep{perrin2009} for scattering angles around 90$^\circ$.

\section{Conclusions}
Our PDI images of the protoplanetary disk around HD169142 reveal previously undetected features and emphasize the power of this observing technique. With PDI on ground-based 8-m class telescopes the inner few tens of AU of the dusty surface layer of protoplanetary disks can be reliably resolved opening up a new era of disk imaging. Several PDI studies in the last few years have shown that sub-structures in disks (e.g., holes, gaps, spiral arms) appear to be rather the rule than the exception \citep[e.g.,][]{quanz2011b, hashimoto2011, muto2012, hashimoto2012,mayama2012,grady2013}. 

In the case of HD169142, our images support previous model results suggesting that the inner $\lesssim$20 AU are largely devoid of scattering dust particles. In addition, our images reveal an annular gap roughy between $\sim$40--70 AU, localized brightness asymmetries on opposite sides of this gap and a discontinuity in the radial surface brightness profile at $\sim$120 AU. Our data alone do not allow us to unambiguously reveal the physical nature of the annular gap, i.e, whether it is a gap in surface density possible induced by forming planet, a shadowing effect, or combination of radially changing dust properties and dust settling. However, our results make HD169142 an excellent target for follow-up studies aiming at a better understanding of disk physics and potentially ongoing planet formation. 

\acknowledgments
This research has made use of the SIMBAD database, operated at CDS, Strasbourg, France. We thank the staff at VLT for their excellent support during the observations and M.R. Meyer, F. Meru and O. Pani{\'c} for useful discussions. This work is supported by the Swiss National Science Foundation.



{\it Facilities:} \facility{VLT:Yepun (NACO)}

\end{document}